\documentclass{PoS}

\usepackage{amsmath,amssymb}
\usepackage{xspace}
\usepackage{lineno}
\newcommand{\eqn}[1]{Eq.~\eqref{#1}}
\newcommand{\GeV}{\ensuremath{\,\mathrm{GeV}}\xspace}
\newcommand{\TeV}{\ensuremath{\,\mathrm{TeV}}\xspace}
\def\eps{\epsilon}
\def\veps{\varepsilon}
\def\mTs{m_t^2}
\def\mHs{m_t^2}

\newcommand{\todo}[1]{}
\title{Next-to-Leading Order Corrections to Higgs Boson Pair Production in Gluon Fusion}

\ShortTitle{NLO Corrections to Higgs Boson Pair Production in Gluon Fusion}

\author{\speaker{Matthias Kerner}%\thanks{A footnote may follow.}
\\
Max Planck Institute for Physics, F\"ohringer Ring 6, 80805 M\"unchen, Germany\\
        E-mail: \email{kerner@mpp.mpg.de}}

\abstract{
We present a calculation of the next-to-leading order QCD corrections to the production of Higgs boson pairs in gluon fusion keeping the full dependence on the mass of the top quark.
The virtual corrections, involving two-loop integrals with up to four mass scales, have been calculated numerically and 
we present an efficient algorithm to obtain accurate results of the virtual amplitude using numerical integrations.
Taking the top quark mass into account we obtain significant differences compared to results obtained in the heavy top limit.
}

\FullConference{Loops and Legs in Quantum Field Theory\\
		24-29 April 2016\\
		Leipzig, Germany}

\begin{document}

\section{Introduction}
%\todo{bbb}
Studying the production of Higgs boson pairs at the LHC is important to scrutinize the mechanism of electroweak symmetry breaking, because this process involves self-interactions of three Higgs bosons and a measurement of this coupling can be directly related to the potential of the Higgs field.
%Studying the production of Higgs boson pairs at the LHC is important to scrutinize the mechanism of electroweak symmetry breaking since the self-interactions of three Higgs bosons appearing in this process can be directly related to the potential of the Higgs field. 

%%%Studying pair production of Higgs bosons at the LHC allows to constraint the self-couplings of the Higgs boson. 
%%%Since this coupling is directly related to the potential of the Higgs field, the measurement of this process is crucial to test the mechanism of electroweak symmetry breaking.

The dominant production mechanism for Higgs boson pairs at the LHC is gluon fusion mediated by a top-quark loop.
The cross section of this process has first been calculated at leading order~(LO) in Ref.~\cite{Glover:1987nx,Plehn:1996wb}.
%,Plehn:1996wb}.
Approximated next-to-leading order~(NLO) results have been obtained in Ref.~\cite{Dawson:1998py} using the Born-improved Higgs Effective Field Theory (HEFT). 
In this approximation the NLO corrections are calculated in the heavy top limit, supplemented with a re-scaling by a factor of $B(m_t)/B(m_t\rightarrow \infty)$, where B is the LO result. Using this approximation also NNLO~\cite{deFlorian:2013uza,deFlorian:2013jea,Grigo:2014jma} and resummed~\cite{Shao:2013bz,deFlorian:2015moa} results have been calculated recently.

In contrast to single Higgs production, it is expected that the heavy top-quark approximation gives only a poor description of Higgs boson pair production since this process peaks in phase space regions where the top-quark mass is not the largest energy scale. Therefore, different methods to improve on the HEFT result emerged during the last few years.
In Refs.~\cite{Frederix:2014hta,Maltoni:2014eza} the $m_t\rightarrow \infty$ limit has only been used to calculate the virtual corrections, whereas the full dependence on $m_t$ has been kept in the real radiation contribution. Another approach has been applied in Refs.~\cite{Grigo:2013rya,Grigo:2015dia,Degrassi:2016vss}, where an expansion in $1/m_t^2$ has been used to improve the prediction for the cross section. In Ref.~\cite{Grigo:2015dia}, see also the presentation in Ref.~\cite{Hoff:2016bdo}, 
this expansion has also been used to improve the predictions for the NNLO contributions.
%is not only applied to obtain the NLO result keeping terms up to $1/m_t^{12}$, but the same method is also applied to the NNLO result keeping terms up to $1/m_t^4$.
%
These two methods for improvements on the HEFT result indicated that the top-quark mass effects should give a contribution of $\mathcal O(10\%)$ at NLO.

In this talk we present the NLO corrections obtained in Ref.~\cite{Borowka:2016ehy}, where the full dependence on the top-quark mass has been kept throughout the calculation. The two-loop integrals appearing in the virtual corrections have been calculated numerically using an interface to the program {\sc SecDec}~\cite{Borowka:2015mxa}.

This calculation is a first step towards the construction of an automated tool for the calculation of general multi-loop amplitudes, based on the programs {\sc GoSam}~\cite{Cullen:2011ac,Cullen:2014yla} and {\sc SecDec}. 
The progress on these developments along with details on the tools used for the calculation have been presented in the talk of S.~Jones~\cite{Stephen}. In the following, we therefore give only a brief overview of these tools and we focus on details of the numerical evaluation of the virtual amplitude and on the phenomenological results.

%\begin{table}[h]
%\begin{tabular}{llrrrrrr}
%  int& num. & s& e & value & error & time [s] & n\\ \hline
%\ldots  &&&\\
%$I_1$ &1&0&all&(0.484, 4.96e-05)&(4.40e-05, 4.23e-05)&11.8459\\
%\ldots&&&\\
%$I_2$ &$(k_1 p_2)(k_2 p_2)$ &0&all&(0.0929, -0.224)&(6.32e-05, 5.93e-05)&235.412\\
%$I_2$ &1&0&all&(-0.0282, 0.179)&(8.01e-05, 9.18e-05)&265.896\\
%$I_2$ & $k_1\cdot p_2\, k_1\cdot p_2$ &0&all&(0.0245, 0.0888)&(5.06e-05, 5.31e-05)&282.794\\
%$I_2$ & $k_1\cdot p_2$ &0&all&(-0.00692, -0.108)&(3.05e-05, 3.05e-05)&433.342\\
%\hline
%&&&  5 & (-1.34e-03, 2.00e-07) & (2.38e-07, 2.69e-07) & 0.255 & 1310420 \\
%&&&6 & (-1.58e-03, -9.23e-05) & (7.44e-07, 5.34e-07) & 0.266 & 1310420 \\
%\ldots &&&&\\
%&&&41 & (0.179, -0.856) & (1.10e-05, 1.22e-05) & 29.484 & 79952820 \\
%&&&42 & (0.359, -1.308) & (1.40e-06, 1.58e-06) & 80.24 & 211436900 \\
%&&&44 & (0.0752, -1.185) & (5.44e-07, 6.76e-07) & 99.301 & 282904860 \\
%
%\end{tabular}
%\end{table}

\section{Calculation}

\begin{figure}[htb]
  \centering
\includegraphics[width=0.9\textwidth]{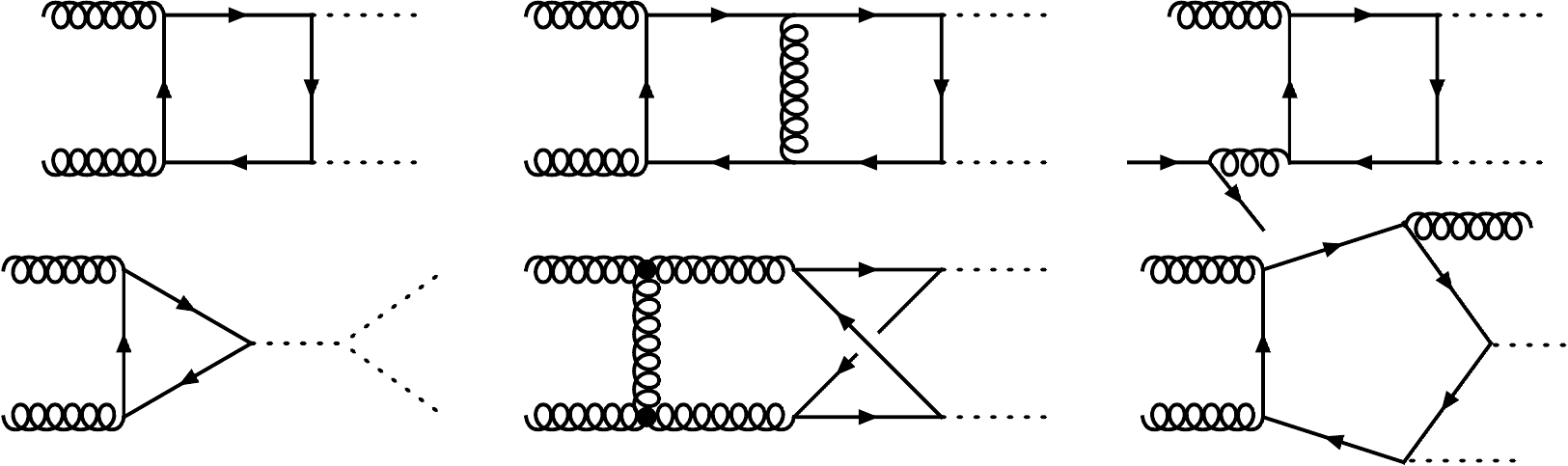}
\caption{Representative Feynman diagrams for the LO, virtual, and real emission contributions.}
\label{fig:Feyn}
\end{figure}
%In the following, we describe the computational setup used to calculate the NLO corrections to Higgs boson pair production. Large parts of this calculation are based 
%
%In the following we 
%Fig.~\ref{fig:Feyn} shows representative Feynman diagrams appearing in the calculation of the NLO corrections to Higgs boson pair production. In the following, we will first describe the general method used for calculating the NLO corrections and we will then describe in detail how the virtual amplitude can be constructed when calculating the loop-integrals numerically.

\subsection{Computational Setup}
Representative Feynman diagrams contributing Higgs boson pair production in gluon fusion are shown in Fig.~\ref{fig:Feyn}.
The amplitude of the underlying $2\rightarrow2$ process 
\begin{align}
  g(p_1) + g(p_2)\rightarrow h(p_3) +h(p_4)
\end{align}
can be decomposed into two form factors $F_1,\,F_2$ as
\begin{align}
&{\cal M}_{ab}=\delta_{ab}\eps_1^\mu\eps_2^\nu\,{\cal
  M}_{\mu\nu}
 % \\
,\quad
&{\cal M}^{\mu\nu}=F_1(s,t,m_h^2,m_t^2,d)\;T_1 ^{\mu\nu}+F_2(s,t,m_h^2,m_t^2,d)\;T_2 ^{\mu\nu}\;,
\label{eq:FFdeco}
\end{align}
where $\eps_1^\mu,\,\eps_2^\nu$ are the polarization vectors of the gluons with color indices $a,\,b$, and with the Mandelstam variables $s=(p_1+p_2)^2$, $t=(p_3-p_1)^2$ and $u=(p_3-p_2)^2$.
The two tensor structures are
\begin{align}
&T_1 ^{\mu\nu}= g^{\mu\nu}-\frac{p_1^{\nu}\,p_2^{\mu}}{p_1\cdot  p_2} \;,\label{eq:Ttensors} \\
&T_2 ^{\mu\nu}= g^{\mu\nu}+
\frac{
  m_h^2 \,p_1^{\nu}\,p_2^{\mu} - 2\,(p_1\cdot p_3) \,p_3^{\nu}\,p_2^{\mu}- 2\,(p_2\cdot p_3) \,p_3^{\mu}\,p_1^{\nu} + 2\,(p_1\cdot p_2) \,p_3^{\nu}\,p_3^{\mu}
}{p_T^2\,(p_1\cdot p_2)}
\end{align}
with $p_T^2=(tu - m_h^4)/s$.
%\begin{align}
%  s&=(p_1+p_2)^2,& t&=(p_3-p_1)^2, & p_T^2&=\frac{tu - m_h^4}{s}.
%\end{align}
%
To calculate the virtual amplitude, we constructed a multi-loop extension of the program {\sc GoSam}
which, after generating the Feynman diagrams using {\sc Qgraf}~\cite{Nogueira:1991ex}, uses {\sc Form}~\cite{Vermaseren:2000nd,Kuipers:2012rf} to apply projectors onto the two form factors  and for further processing of the expressions. 
To facilitate the use of reduction programs, we created an interface to {\sc Reduze}~\cite{vonManteuffel:2012np} and matched all loop integrals to integral families, leading to expressions for $F_1$ and $F_2$ containing $\sim$10.000 integrals with up to 7 propagators and 4 inverse propagators. These expressions have been validated by comparing them to a second implementation entirely based on {\sc Qgraf} and {\sc Reduze}.

The reduction of the integrals to master integrals turned out to be very challenging and we therefore fixed the numeric values $m_t=173\GeV$ and $m_h=125\GeV$ during reduction, thus reducing the number of appearing mass scales by one. With this simplification, the reduction to masters has been obtained for all the planar integrals, but we didn't achieve a full reduction for the non-planar 6 and 7 propagator integrals using the programs {\sc Reduze}~\cite{vonManteuffel:2012np}, {\sc Fire}~\cite{Smirnov:2014hma} or {\sc LiteRed}~\cite{Lee:2013mka}. For simplifying the numerical evaluation of these integrals, we therefore rewrote the inverse propagators in terms of scalar products, leading to non-planar 7 propagator integrals of up to rank 4. 

After the partial reduction, the expressions for the amplitude contained 145 planar master integrals as well as 70 mostly unreduced non-planar tensor integrals, leading to a total of 327 integrals when including integrals that differ by a crossing. 
We calculated these integrals using the program {\sc SecDec},
which decomposes the integrals into sectors leading to finite Feynman parameter integrals at each order in $\eps=(4-d)/2$, where $d$ is the dimension of space-time.
We modified {\sc SecDec} such that the integrands of latter integrals are written to a library, which allows us to call them directly from our code for the amplitude evaluation and we dynamically adjust the number of sampling points as discussed in the next section.
For the numerical integration we use a quasi-Monte Carlo algorithm based on a rank-one lattice rule~\cite{QMCActaNumerica,Li:2015foa}, 
which  for sufficiently smooth integrands  obeys a scaling  $\Delta\propto n^{-1}$  of the integration error with the number of sampling points.
%This allows us to obtain accurate results for the individual integrations.
%The sampling of phase-space points is done using unweighted events generated based on the LO result.
\todo{todo:finite basis?, unweighted events?}

The real emission contribution involves up to one-loop pentagon diagrams with a top-quark appearing in the loop. The corresponding amplitudes
have been generated using the program {\sc GoSam}
%, which allows for automated generation of one-loop amplitudes
and we used the dipole subtraction algorithm~\cite{Catani:1996vz}
to deal with the infrared singularities.
%and we restrict the phase-space of the subtraction terms using the method presented in Ref.~\cite{Nagy:2003tz}.
%The LO matrix elements required for the subtraction terms were calculated using the analytic results given in Ref.~\cite{Glover:1987nx} and the phase space integration has been performed using {\sc Cuba}~\cite{Hahn:2004fe}.
\todo{todo: mention alpha parameter, Cuba?}

\subsection{Numerical Evaluation of the Virtual Amplitude}
After form factor decomposition and partial reduction, the virtual amplitude $\mathcal M_V$ is determined by the two form factors $F_1$ and $F_2$, which are a linear combination of 327 two-loop integrals $I_j$,
\begin{align}
  F_i(s,t,\mTs,\mHs,\eps) = \sum_j f_{i,j}(s,t,\mTs,\mHs,\eps)\cdot I_j(s,t,\mTs,\mHs,\eps),
  \label{eq:FFints}
\end{align}
with multi-variate rational functions $f_{i,j}$ as coefficients. For the numeric calculation of the integrals, it is convenient to factor out an arbitrary mass scale $M$ as well as a prefactor $c_j(\eps)$ containing the $\Gamma$-functions arising from Feynman parametrization and integration of the loop momenta. This leads to
\begin{align}
  I_j(s,t,\mTs,\mHs;\eps) = \left( \frac{\mu^2}{M^2} \right)^{2\eps} c_j(\eps)\, M^{m_j}\, \hat I_j\left(\frac{s}{M^2},\frac{t}{M^2},\frac{\mTs}{M^2},\frac{\mHs}{M^2} ;\eps \right),
\end{align}
where $\mu$ is the mass scale of dimensional regularization and the exponent $m_j$ is given by the mass dimension of integral $I_j$.
After applying sector decomposition using {\sc SecDec}, each loop integral $\hat I_j$ is decomposed into multiple sectors $s$, which can be expanded in $\eps$, leading to a Laurent series
\begin{align}
  \hat I_j
\left(\frac{s}{M^2},\frac{t}{M^2},\frac{\mTs}{M^2},\frac{\mHs}{M^2} ;\eps \right)
%  (s,t,\mTs,\mHs;\eps)
  =\sum_s\, \sum_{e>e_s^{\mathrm {min}}}\, \eps^e\,\hat I_{j,s,e}%(s,t,\mTs,\mHs).
\left(\frac{s}{M^2},\frac{t}{M^2},\frac{\mTs}{M^2},\frac{\mHs}{M^2}  \right),
\label{eq:secdec}
\end{align}
and the integrals $\hat I_{j,s,e}$ can be integrated numerically over the Feynman parameter space.
%$e_s^{\mathrm{min}}$ is given by the singularity structure of sector $s$ and $c_j(\eps)$ is the prefactor containing the $\Gamma$-functions stemming from Feynman parametrization and integration over the loop momenta.
Rearranging the terms of Eqs.~\eqref{eq:FFints}-\eqref{eq:secdec}, we can write the two form factors as
\begin{align}
  F_i(\eps) = \left( \frac{\mu^2}{M^2} \right)^{2\eps} \cdot \underbrace{\sum_{j,s,e} \hat I_{j,s,e}\cdot a_{i,j,e}(\eps)}_{A_i}
  \label{eq:FFfinal}
\end{align}
and expand the coefficients
\begin{align}
%  a_{i,j,e}\left(\frac{s}{M^2},\frac{t}{M^2},\frac{\mTs}{M^2},\frac{\mHs}{M^2};\eps  \right)
%  = \eps^e M^{m_j} c_j(\eps) f_{i,j}\left(\frac{s}{M^2},\frac{t}{M^2},\frac{\mTs}{M^2},\frac{\mHs}{M^2};\eps  \right)
  a_{i,j,e}(\eps) = \eps^e M^{m_j} c_j(\eps) f_{i,j}(\eps)
\end{align}
up to $\mathcal O(\eps^0)$, where the dependence of the form factor, integral and coefficients on $s,t,\mTs$ and $\mHs$ has been suppressed. 
This form allows us to obtain the results of the two form factors (including their poles in $\eps$), while computing each integral only once. Furthermore, applying a similar procedure to the calculation of the LO amplitude and mass counter terms, \eqn{eq:FFfinal} allows us to vary the renormalization scale $\mu_R$ without recomputing  $A_i$.

Sector decomposition allows us to write the amplitude in terms of finite integrals which we can calculate numerically, however, it also leads to a significant increase in the number of integrals. After decomposition and expansion in $\eps$, the 327 integrals $I_j$ are replaced by $\sim$11.000 integrals $\hat I_{j,s,e}$.
With this large number of integrals, it is not advisable to evaluate each integral with a pre-defined accuracy or number of sampling points.
%The numerical integration of all integrals with a pre-defined number of points or precision goal would lead to several problems: 
Assuming that each integral gave the same contribution to the amplitude, each integral would be required with a precision of $\mathcal O(10^{-4})$ to obtain an amplitude result with an accuracy of 1\%.
However, cancellations between various integrals can spoil the accuracy of the amplitude result and a significant increase in the precision for these integrals might be required. 
On the other hand, one wants to avoid calculating integrals to very high accuracy if they only give a small contribution. 

 We therefore dynamically set the number of sampling points for each integral depending on its contribution to the error estimate of the amplitude result and depending on the time required for each integrand evaluation. 
 For simplicity, we now consider the evaluation of only one form factor at a given power in $\eps$ and we write \eqn{eq:FFfinal} as $F=\sum_j (a_j\,I_j)$. 
 To efficiently calculate $F$ to a given accuracy $\veps_{\mathrm{rel}}=\Delta_F/F$, we set the number of sampling points for each integral by minimizing the total time 
 \begin{align}
   T=\sum_j t_j + \lambda \left( \Delta_F^2-\sum_j a_j^2 \Delta_j^2 \right),
   \label{eq:opt}
 \end{align}
 where $t_j$ and $\Delta_j$ are the integration time and absolute error of integral $I_j$, and 
 $\lambda$ is a Lagrange multiplier ensuring the accuracy constraint.
 After an initial run of each integral with a fixed number of sampling points, we set the number of sampling points according to \eqn{eq:opt} assuming that the error of the individual integrals 
 $I_j$ scales as $\Delta_j \propto t_j^{-e}$. While we obtain a scaling with $e=1$ for most of the integrals, some of the integrals don't fulfill the smoothness condition of the quasi-Monte Carlo method leading to worse convergence of these integrals and we therefore set $e=0.7$ globally in the program.
\todo{todo: improve last paragraph, maxincrease, different Feps}  
\todo{MC error reliable}

\section{Phenomenological Results}
We present results for the cross section of Higgs boson pair production at the LHC with a center of mass energy of $\sqrt s=14 \TeV$.
The masses of the Higgs boson and top quark are set  to $m_h=125$\,GeV and $m_t=173$\,GeV, and 
we use the PDF4LHC15\_nlo\_100\_pdfas~\cite{Butterworth:2015oua,CT14,MMHT14,NNPDF} parton distribution functions,
along with the corresponding value for $\alpha_s$. The central value of the renormalization and factorization scale are set to $\mu_R=\mu_F=m_{HH}/2$ and we estimate the scale uncertainty by simultaneously varying these scales by a factor of 2.

We obtain a total cross section of
\begin{equation*}
  \sigma^{NLO} = 32.90^{+14\%}_{-13\%} 
  \,\mathrm{fb}
  \pm 0.3\%\, (\mathrm{stat.}) % (=0.104 fb) 
  \pm 0.1\%\, (\mathrm{int.}). % (=0.03 fb),
\end{equation*}
where we state the statistical error stemming from the number of evaluated phase points and the additional error due to the numerical integration of the virtual amplitude in addition to the scale uncertainty. This result is a factor of $\sim$1.6 larger than the LO result $\sigma^{LO}=19.85^{+28\%}_{-21\%}$ and 14\% smaller than the result $\sigma^{\mathrm{NLO}}_{\mathrm{HEFT}}=38.32^{+18\%}_{-15\%}$ obtained in the Born-improved HEFT approximation.

\begin{figure}[tb]
\includegraphics[width=0.5\textwidth]{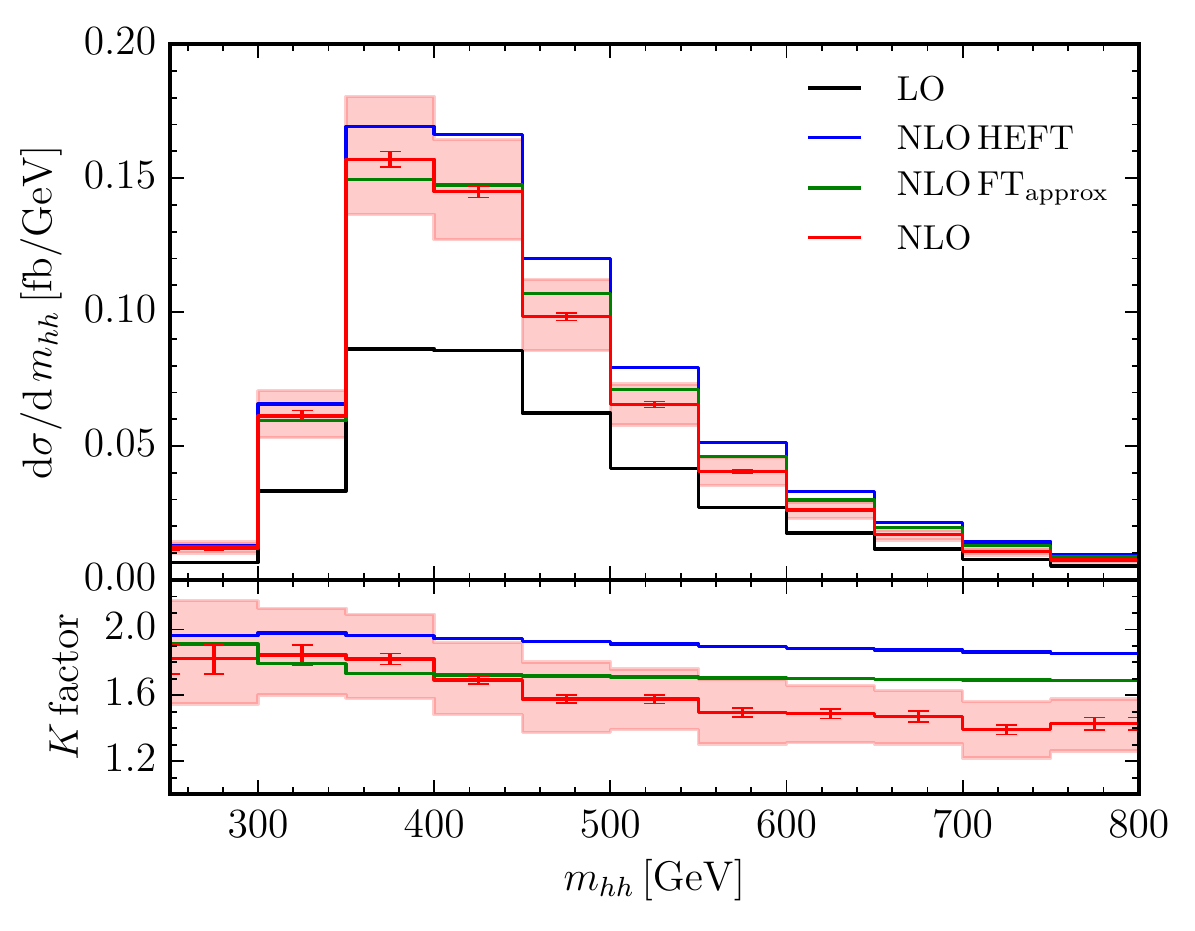}
\includegraphics[width=0.5\textwidth]{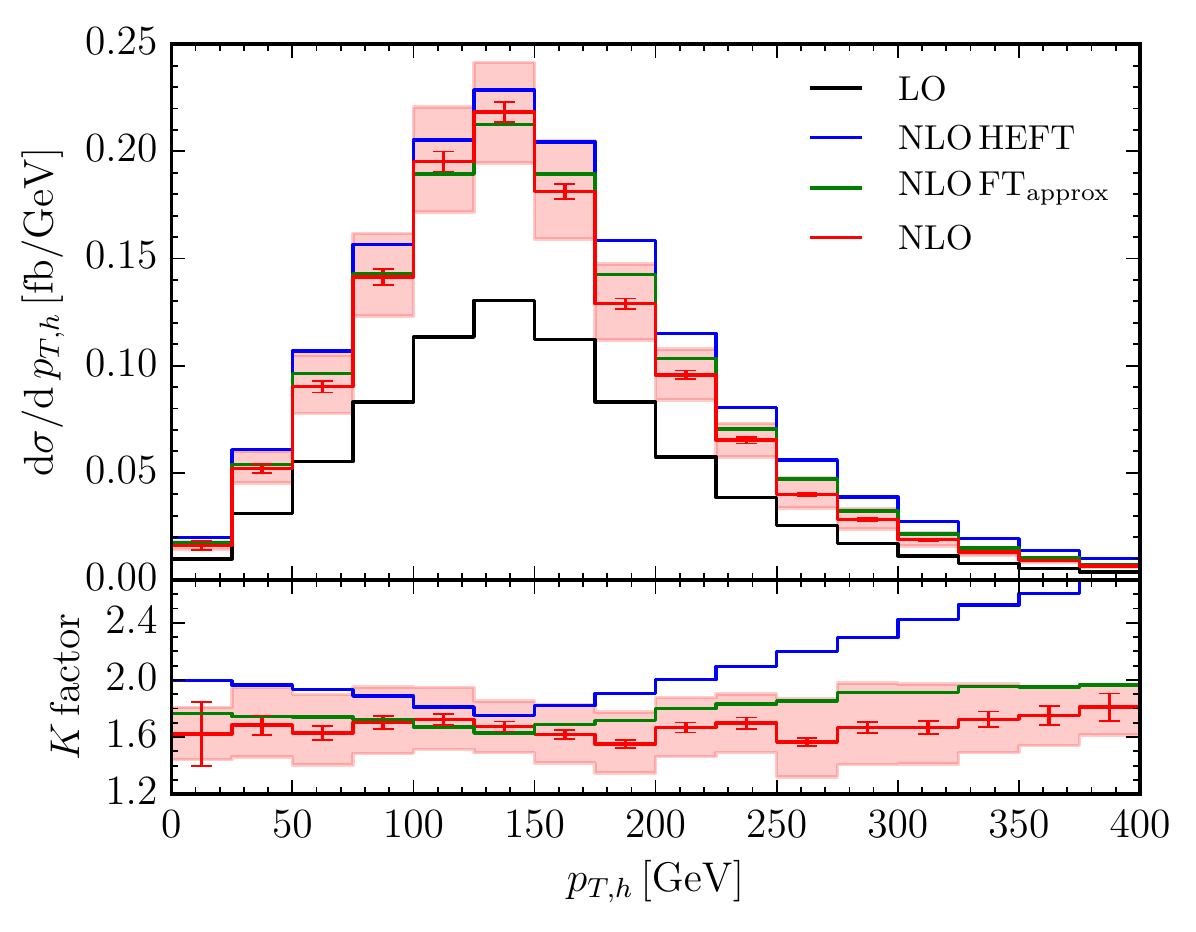}
\caption{
  Differential distributions of the invariant mass $m_{hh}$ of the Higgs boson pair and transverse momentum $p_{T,h}$ of the Higgs bosons. 
The  center of mass energy is set to $\sqrt{s}=14$\,TeV and the bands result from scale
variations by a factor of two around the central scale $\mu=m_{hh}/2$.
  We compare our predictions to the approximated NLO results using the Born-improved HEFT and taking the top-quark mass effects into account only in the real emission contributions, FT$_{approx}$.
\label{fig:fullresult}}
\end{figure}

The differential dependence of the cross section on the invariant mass $m_{hh}$ of the Higgs bosons as well as the dependence on their transverse momentum $p_{T,h}$ is shown in Fig.~\ref{fig:fullresult}. We find good agreement of our predictions with the approximated NLO results for invariant masses below 400\GeV, where the HEFT is expected to be valid. However, for large invariant masses we observe that taking the top-quark mass effects into account reduces the cross section by about 20-30\% compared to the HEFT result. In the $p_{T,h}$-distribution we obtain a nearly constant K factor, whereas the HEFT approximation leads to a significant increase of the NLO corrections for high transverse momenta.  Compared to the HEFT results, the NLO FT$_{approx}$ results, which include the top-quark mass effects in the real emission, lead to better agreement with the full NLO result. However, also this approximation fails to describe the decreasing K factor for high invariant masses.

\begin{figure}[tb]
  \centering
  \includegraphics[width=0.5\textwidth]{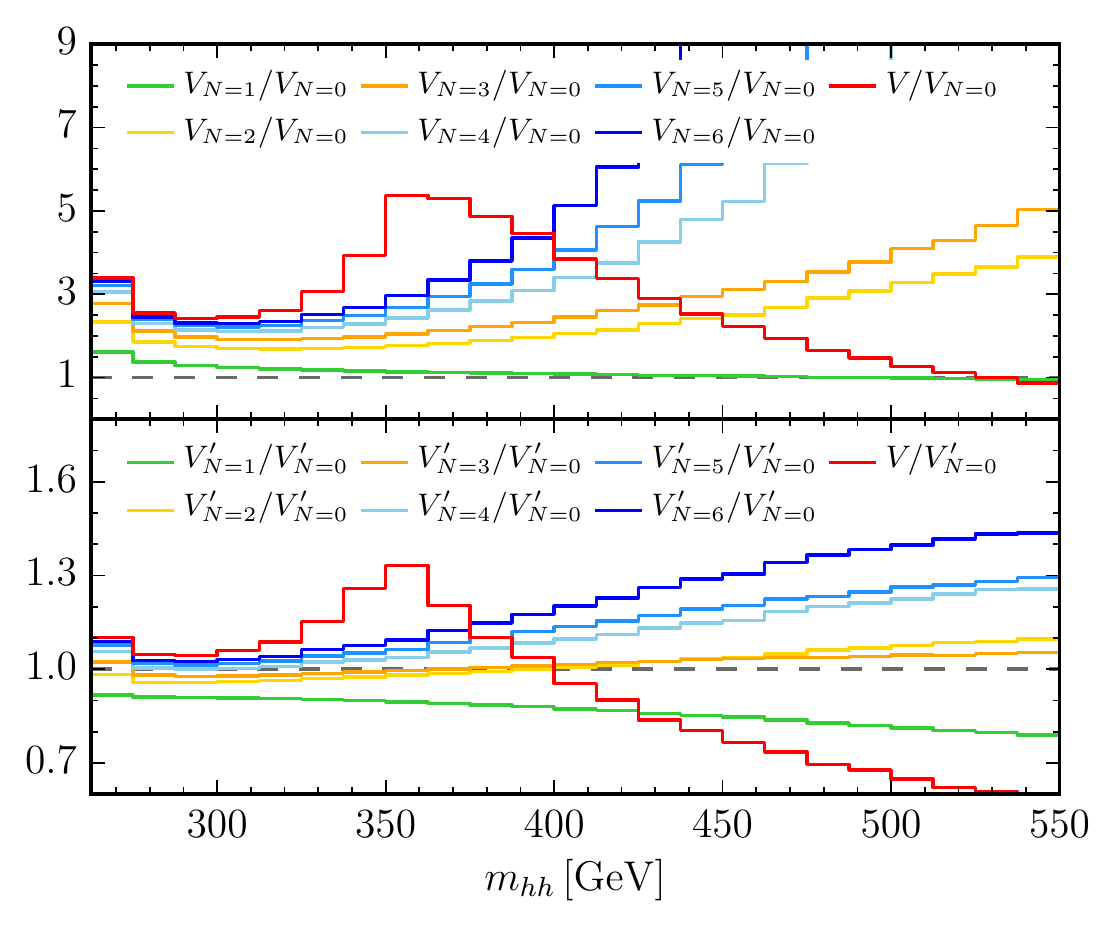}
  \caption{
    Comparison of the virtual contributions $V$ with the results obtained by an expansion in $1/m_t^2$. Results are shown at the amplitude level with the definitions of  $V_N$ and $V'_N$ given in
  Eq.~\protect\eqref{eq:Vexp}.
  }
\label{fig:Vexp}
\end{figure}

Fig.~\ref{fig:Vexp} shows a comparison of the virtual amplitude with an expansion in $1/m_t^2$ obtained from  private communication with the authors of Ref.~\cite{Grigo:2015dia}. We combine the renormalized amplitude with the $\mathbf I$-operator defined in Ref.~\cite{Catani:1996vz} and define
\begin{align}
  V_N=d\sigma^{\mathrm{virt}}_N + d\sigma^{\mathrm{LO}}_N\otimes \mathbf I  \quad \mathrm{and} \quad V'_N=V_N\cdot \frac{d\sigma^{\mathrm{LO}}}{d\sigma^{\mathrm{LO}}_N}
  \label{eq:Vexp}
\end{align}
where  $d\sigma^{LO}_N$ and $d\sigma^V_N$ are the LO and virtual contribution expanded up to order $1/m_t^{2N}$. 
The figure shows that the expansion converges to the full virtual amplitude in the region well below the top-quark threshold at $m_{hh}=2m_t$ but we obtain large differences at higher invariant masses.

\section{Conclusions}
We presented a computation of the NLO corrections to Higgs boson pair production in gluon fusion, keeping the full dependence on the top-quark mass throughout the calculation.
Since the results for the two-loop integrals appearing in the virtual amplitude are not known analytically, we calculate these numerically and we have shown how an accurate result for the amplitude can be obtained using this numerical approach. 
Using the methods presented here, we plan to develop a program for the automated generation and evaluation of multi-loop amplitudes.

Our predictions show that the effects of the top-quark mass are important and they have to be taken into account to obtain reliable predictions for Higgs boson pair production in gluon fusion. 
For low invariant masses of the di-Higgs system we find good agreement of the HEFT result with our predictions. However, we observe that the inclusion of the top-quark mass reduces the cross section by $\sim$30\% in phase space regions where $m_{hh}$ or $p_{T,h}$ is large. For the total cross section at 14\TeV we obtain a reduction of 14\% compared to the result obtained the Born-improved HEFT approximation.

%\section*{Acknowledgment}
\acknowledgments
I am grateful to Sophia Borowka, Nicolas Greiner, Gudrun Heinrich, Stephen Jones, Johannes Schlenk, Ulrich Schubert, and Tom Zirke for the fruitful collaboration during this and ongoing projects. Furthermore, I want to thank Jens Hoff for providing us with the expansion of the virtual amplitude in $1/m_t^2$.

\bibliographystyle{JHEP}
\bibliography{refsHH}
%\begin{thebibliography}{99}
%\bibitem{...}
%....
%
%\end{thebibliography}

\end{document}